\documentclass[11pt]{article}
\usepackage{latexsym}
\usepackage{graphics}

\newcommand{\be}{\begin{equation}}
\newcommand{\eq}{\end{equation}}
\def\square{\sqcup\!\!\!\!\sqcap}

\newcommand{\Sum}{\displaystyle\sum}

\newcommand{\R}{{\rm I\!R}}
\newcommand{\Z}{Z\!\!\!Z}
\def\un { \rm 1\!\!I}
\newcommand{\finprf}{\unskip\null\hfill$\square$\vskip 0.3cm}

\newcommand{\affil}[1]{{\small\sl #1}}
\newcommand{\email}[1]{{\small E-mail: {\textsf {#1}}}}
\newcommand{\http}[1]{{\small Internet: {\textsf {#1}}}}

\newtheorem{theorem}{Theorem}
\newtheorem{proposition}[theorem]{Proposition}
\newtheorem{lemma}[theorem]{Lemma}
\newtheorem{corollary}[theorem]{Corollary}

\begin{document}
\title{\sl Erratum : Existence of $\,3D\,$ Skyrmions. Complete version.}
\author{
Maria J. ESTEBAN\\
\affil{Ceremade (UMR CNRS no. 7534), Universit\'e Paris IX-Dauphine,}\\
\affil{Place de Lattre de Tassigny, 75775 Paris C\'edex~16, France}\\
\email{esteban@ceremade.dauphine.fr}\\
\http{http://www.ceremade.dauphine.fr/\raisebox{-4pt}
{$\widetilde{\phantom{x}}$}esteban/}\\
}
\maketitle
\thispagestyle{empty}

\begin{abstract}
This erratum corrects the proof given in \cite{E1,E2} about the existence of $\,3D\,$ Skyrmions.  This is done by changing the arguments of the proof while remaining in the same framework of concentration-compactness. Note however that the use of this method is here different of most of what has been done  with it so far. In that sense, this new proof has some interest by itself. The proof  given here is self-contained. I thank F. Lin and Y. Yang for having pointed out to me that there were gaps in my proofs.

\end{abstract}
\section{Introduction}
\label{intro}

The Skyrme's problem consists in minimizing an energy functional together with a condition, which is of topological type (see the papers in the reference list for information on the physical meaning of this problem). The functional space consists of functions mapping $\,\R^3\,$ into $\,S^3$.
In \cite{E1}  an existence result was proved for Skyrmions in $\,3D\,$ by using the concentration-compactness method.   Then, in \cite{E2}   the same result was proved in another functional context, but using  the same technical arguments. But as Fanghua Lin and Yisong Yang have pointed out to me, the proof of the main result contained in \cite{E1,E2}  is not correct. In this Erratum  a new proof is made
by changing  the arguments used in \cite{E1,E2}. The method used here is still the concentration-compactness principle but applied in a different, and in some sense, less usual way.

In a very interesting paper basically devoted to the study of the Faddeev knots (\cite{LY1}),  F. Lin and Y. Yang prove the existence of $\,3D\,$ Skyrmions of degree $\,\pm1\,$  by using a different approach, which is based on a cubic decomposition of the whole space. In that paper, they obtain a condition for the existence of solutions for the $\,3D\,$ Skyrme's problem  given by a family of strict decomposition inequalities.   After communicating their paper to me, I realized that my previous proof contained some gaps and I tried to correct it in the same spirit of concentration-compactness that I had used from the beginning. By doing this, I reach the same condition for the existence of minimizers. This is not surprising. Indeed, the above family of strict inequalities is not only sufficient for the existence of minimizers, but is in fact  necessary and sufficient for the relative compactness of all minimizing sequences, so it is not so surprising that by different approaches we reach the same condition. 

One of the main problems in the proofs in \cite{E1,E2}  is a cutoff lemma allowing to approach a Skyrme's finite energy function by a function which is constant near infinity. Or separating into two distinct finite energy functions one which is ``almost constant" in a large annulus-like domain. And the possibility of doing this in $3D$ is still open.  In another excellent paper, 
\cite{LY2} F. Lin and Y. Yang show that this can be done in dimension $2$, which is a very nice result. So, after they prove this, they can obtain the existence of $\,2D\,$ Skyrmions by ``more classical" concentration-compactness arguments. In $\,2D\,$ the arguments are somehow different due to the presence of a ``new" term in the energy functional.

Here I prove again the main existence result in \cite{E1,E2} and the proof is self-contained. In order to write a short erratum I quote some auxiliary  results which are proved in \cite {E1,E2,E3} and then,  wholly prove the main theorem.

Let me note again that the way to tackle the possible losses of compactness of minimizing sequences is not exactly the same in this paper and in that of Lin and Yang, even if in the end we reach the same condition for existence of Skyrmions, and hence the same theorem.
\section{ Main theorem and auxiliary results.}
\label{main}
For all functions $\,\phi: \R^3\to S^3\,$ which are of class $\,C^1\,$ and constant outside a ball of $\,\R^3\,$ we can define a notion of topological degree, which is an integer and which can be represented by the expression
\be d(\phi)={1\over{2\,\pi^2}}\int_{\R^3} \det(\phi, \nabla \phi)\,dx\,.\eq

In \cite{E3} we proved that this notion of topological degree can be extended to all functions $\,\phi:\R^3\to S^3\,$ such that $\,\nabla \phi\,$ and $\,\partial_i\phi\wedge \partial_j\phi\,$ ($i, j\in \{1,2,3\}$) belong to $\,L^2(\R^3, S^3)$, this degree still  being an integer.

If we denote by $\,A(\phi):= \Big(\partial_i\phi\wedge \partial_j\phi\Big)\,$ ($i,j=1,2,3$), then the Skyrme's problem consists in minimizing the energy
\be {\mathcal E}(\phi):=\int_{\R^3} |\nabla \phi|^2+|A(\phi)|^2\, dx\,,\eq
over the functions $\,\phi\in X:= \{\phi:\R^3\to S^3\;; \;{\mathcal E}(\phi)<+\infty\}\, $ which have a given degree $\,d(\phi)=k\,$  and we denote this minimum by $\,I_k$. The minimizers are called Skyrmions. Note that the case of degree $0$ is trivial, since then the minimizers are just the constant functions and $\,I_0=0$. So, in what follows we only address the case $\,k\ne 0$.

Here we prove the following 
\begin{theorem}\label{T1} Let $\,k\ne 0$. If for all  finite integer decompositions of $\,k\,$, $k=\sum_{i=1}^J d_i\,$, $d_i\in \Z\setminus\{0\}$, 
\be\label{ineq} I_k<\sum_{i=1}^J I_{d_i}\,,\eq
then, $I_k$ is achieved. \end{theorem}

Note that in \cite{E1,E2} only binary decompositions ($J=2$) had to be avoided. The difference lies in the fact that we do not know anymore whether  for all $\ell\in \Z\setminus\{0, k\}$,  the large inequalities $I_k\leq I_\ell+I_{k-\ell}\,$  hold or not. 

\smallskip

Moreover, as in \cite{E0}, we have
\begin{proposition}\label{estimate} For  all integer $\,k\,$,
\be\label{estt} 12\,|k|\,\pi^2\leq I_k \leq 12\,\sqrt{2}\,|k|\,\pi^2\,.\eq
\end{proposition}

Hence, from Theorem \ref{T1} and the above Proposition  we obtain
\begin{corollary}  The two infima $\,I_{\pm 1}\,$ are achieved, that is, there exist minimizing Skyrmions with degree $\,\pm1$.\end{corollary}

\section{Proof of main results.}\label{proof}

\noindent{\bf Proof of Proposition \ref{estimate}}. The first inequality in (\ref{estt}) follows easily from Schwartz and H\"older inequalities: For all $\,\phi\in X$,
$${\mathcal E}(\phi)\geq 2 \left(\int_{\R^3} |\nabla \phi|^2\, dx\right)^{1/2} \left(\int_{\R^3} |A(\phi)|^2\, dx\right)^{1/2}$$ $$\geq 6 \int_{\R^3} |\partial_1\phi\wedge\partial_2\phi\wedge\partial_3\phi|\, dx\geq 12\pi^2\,|d(\phi)|\,.$$

On the other hand, if we consider the stereographic projection from $\,S^3\,$ into $\,\R^3$, its inverse $\,\tilde\phi\,$ belongs to $\,X\,$ and its energy can be computed easily and it is equal to $\,12\sqrt{2}\,\pi^2$ (see \cite{E0}). Moreover, $\,\tilde\phi\,$ can be approximated by functions $\,\tilde\phi_n\in X$ which are constant outside a big ball and such that 
$$\lim_{n\to +\infty}\,{\mathcal E}(\tilde\phi_n)=12\sqrt{2}\,\pi^2\,.$$ 
One can ``link" $\,k\,$ copies (maybe conveniently rotated)  of those $\,\tilde\phi_n$'s to construct a map of degree $\,k\,$ with energy as close  to $\,12\sqrt{2}\,|k|\,\pi^2\,$ as desired. \finprf

In order to prove our main result, let us begin by noting that
 by the Poincar\'e-Wirtinger inequality, it is easy to see that there is a constant $\,C>0\,$ such that for all $\phi\in X$, up to a rotation, 
\be\label{pw}\int_{\R^3}|\phi-P|^6\,dx< C\left(\int_{\R^3}|\nabla\phi|^2\,dx\right)^3\,,\eq
where $P$ is the north pole of the sphere $S^3$ (see \cite{E1}, Lemma 7). Note that the problem is invariant by rotation and translation.

Two important auxiliary results for the proof of Theorem \ref{T1} are the following propositions.

\begin{proposition}\label{3/4}(\cite{E1})
For any $\,\phi\in X$, for any $\,B\subset\R^3\,$ measurable,
\be\int_{B}\left|\partial_1\phi\wedge\partial_2\phi\wedge\partial_3\phi\right| \,dx \leq |B|^{1/4}
\left(\int_{B}|A(\phi)|^2\, dx\right)^{3/4}\,.\eq
\end{proposition}

The proof of this proposition can be found in \cite{E1}.

\begin{proposition}\label{degre0} Let $\,\phi\in X$. Then, $P$ being the north pole of $\,S^3$, 
\be\label{Pseul}\int_{\R^3}\det(P, \nabla\phi)\,dx=0\,.\eq
\end{proposition}

\noindent{\bf Proof.}  Let $\,\varphi\in{\mathcal D}(\R^3, \R)^4\,$ and $\,\phi\in X$. Then, 
\be\label{luze}\int_{\R^3} \!\!(\varphi\wedge\partial_1\phi\wedge\partial_2\phi\wedge\partial_3\phi)\,dx=
<\!\varphi\wedge\phi, P(\phi) \!>\! -\!\int_{\R^3} (\partial_1\varphi\wedge\phi\wedge\partial_2\phi\wedge\partial_3\phi)\,dx\eq
$$-\int_{\R^3} (\partial_2\varphi\wedge\partial_1\phi\wedge\phi\wedge\partial_3\phi)\,dx
-\int_{\R^3} (\partial_3\varphi\wedge\partial_1\phi\wedge\partial_2\phi\wedge\phi)\,dx
\,,$$
where in this identity $\,<\cdot, \cdot>\,$ stands for the duality product between $\,{\mathcal D}^{1,2}\,$ and its dual, and 
$$P(\phi):=\partial_2(\partial_1\phi\wedge\partial_3\phi)-
\partial_1(\partial_2\phi\wedge\partial_3\phi)-\partial_3(\partial_1\phi\wedge\partial_2\phi)\,.$$

Now let $\,\{\rho_\epsilon\}_\epsilon\subset{\mathcal D}(\R^3)\,$ a regularizing sequence such that if for 
$\,\phi\in X$, we define $\,\phi_\epsilon:=\phi*\rho_\epsilon:\R^3\to \R^4$, we have $\displaystyle{\nabla\phi_\epsilon\longrightarrow_{_{\hskip-5mm\epsilon\to 0}} \nabla\phi}$ in $\,L^2(\R^3)$. As a consequence, $\,P(\phi_\epsilon)\,$ converges towards $\,P(\phi)\,$ in the distributional sense as $\epsilon$ 
goes to $0$. On the other hand, an easy computation shows that for every $\epsilon>0$, $\,P(\phi_\epsilon)=0$. Hence, the distribution $\,P(\phi)\,$ is the null distribution and therefore the term $\,<\varphi\wedge\phi, P(\phi) >\,$ in the r.h.s. of (\ref{luze}) is equal to $0$.

Now, take a sequence $\,\{\varphi_n\}_n\,$ in $\,{\mathcal D}(\R^3, \R)^4\,$, bounded in $W^{1,\infty}(\R^3, \R)^4\,$ and such that $\,\varphi_n\,$ converges towards $P$ a.e. as $n$ goes to $+\infty$. Then, the proposition is proved by writting (\ref{luze}) for $\,\varphi=\varphi_n\,$ and passing to the limit in $n$ with the help of  Lebesgue's Theorem.  \finprf

We can start now the 

\medskip
\noindent{\bf Proof of Theorem \ref{T1}}.
Let $\,\{\phi_n\}\,$ be a minimizing sequence for $\,I_k$. We define
$$f_n:= |\nabla\phi_n|^2+|A(\phi_n)|^2+|\phi_n-P|^6\,.$$
Up to subsequences we may assume that $\displaystyle{\,\lim_{\,n\to+\infty}\,\int_{\R^3}f_n\, dx =A}$,  $A\in ( I_k, +\infty)$.

By (\ref{pw}), $\,\{f_n\}_n\,$ is a bounded sequence of nonnegative functions in $\,L^1(\R^3)$. Hence, we can apply to it the first concentration-compactness Lemma of P.-L. Lions \cite{PLL1}, which states that up to extraction of subsequences, either
\begin{itemize}
\item (vanishing) for all $R>0$, $\displaystyle{\,\lim_{n\to +\infty}\,\sup_{y\in \R^3}\int_{B(y,R)  }f_n\,dx=0\,,}$ or
\item (dichotomy) there exist a sequence $\,\{y_n\}_n\,$ in $\,\R^3\,$ and numbers $\,a_1, b_1>0\,$ such that $\,a_1+b_1=A\,$   and  for all $\epsilon>0$ there exist $\,R>0\,$ and a sequence  of positive numbers $\,\{R_n\}_n\,$  such that $\,R_n\to_n +\infty$ and 
\be \left|\,a_1-\int_{B(y_n,R)  }f_n\,dx\, \right|\leq \epsilon\;, \quad \left|\, b_1-\int_{\R^3\setminus B(y_n,R_n)  }f_n\,dx \, \right|\leq \epsilon\,, \eq
and $\,a_1\,$ is the maximal number satisfying this property, or
\item (compactness) there exists a sequence $\,\{y_n\}_n\,$ in $\,\R^3\,$ such that  for all $\epsilon>0$ there exists $\,R>0\,$ with 
\be \int_{\R^3\setminus B(y_n,R)  }f_n\,dx \leq \epsilon\,,\quad\mbox{for all }\, n\,. \eq

\end{itemize}

To deal with the possibility of vanishing we prove the following 
\begin{lemma}\label{vanish} Let $\,\{D_n\}_n\,$ be  a sequence of  measurable subsets of $\,\R^3\,$.  If there is vanishing for the sequence $\,\{f_n\,\un_{D_n}\}$, then for $\,n\,$ large enough,
\be\lim_{n\to +\infty}  \int_{D_n} \det(\phi_n-P, \nabla \phi_n)\,dx=0\,\,. \eq
\end{lemma}

\noindent{\bf Proof.}
Let us define
$$ c_n:=\sup_{y\in \R^3}\int_{B(y,1) \cap D_n }f_n\,dx\,.$$

Since the sequence $\,\{\phi_n\}\,$ is uniformy bounded in $\,L^\infty\,$ and since in three dimensions  $W^{1,1}(B(y,1))$ is embedded in $\,L^1(B(y,1))\cap L^{3/2}(B(y,1))\,$, the embedding constant being independent of $\,y$, then for all $\,\alpha\in (1, 3/2)$ and for all $\,y\in \R^3$,
$$\int_{B(y,1)}|\phi_n-P|^{6\alpha}\,dx\leq C\left(\int_{B(y,1)}|\phi_n-P|^6+|\nabla \phi_n|^2\,dx\right)^\alpha\,,$$
$C$ being independent of $n$ and of $\,y$. Considering now a locally finite covering of  $\,\R^3$, $\,\{B(y_i,1)\}_{i\geq 1}\,$ such that every point of $\,\R^3\,$ is at most in $\,m\,$  of those balls, we have
\begin{equation}\label{van}\int_{D_n}\left(|\phi_n-P|^{6\alpha}\right)\,dx\leq C\,m\,c_{\,n}^{\,\alpha-1}\int_{\R^3}|\phi_n-P|^6+|\nabla \phi_n|^2\,dx\,.\end{equation}

Hence, choosing $\,\alpha=7/6\,$ we get $\, \lim_{n\to +\infty}||\phi_n-P||_{L^7(D_n)}=0$.  Now define
$$A_{n}:=\{x\in  D_n\;;\quad |\phi_n-P|\geq c_n^\gamma \quad\hbox{a. e.}\}\,,$$
with $\,\gamma>0\,$ such that $\,\frac{1}6-7\gamma>0$.

By Proposition \ref{3/4} and the hypothesis of vanishing, we can find $\,C'>0$ independent of $n$ and  of $\epsilon$ such that
$$\left|\int_{D_n\setminus A_{n}}\!\!\!\!\!\!\det(\phi_n-P, \nabla \phi_n)\, dx\right| \leq C' c_n^\gamma\,,\;\;
\left|\int_{A_{n}}\!\!\!\! \det(\phi_n-P, \nabla \phi_n)\, dx\right| \leq C' |A_{n}|^{1/4}\,.$$
while for $n$ large enough,  (\ref{van}) implies that  $|A_{n}|\leq C'\,c_n^{\,\,\frac{1}6-7\gamma}$. Hence, the result.
 \finprf

 Lemma \ref{vanish} applied with $\,D_n=\R^3\,$ forbids vanishing for any minimizing sequence of $\,I_k\,$ as soon as $\,k\ne 0$. Indeed, use Proposition \ref{degre0} to infer that $$\int_{\R^3} \det(\phi_n-P, \nabla\phi_n)\,dx= \int_{\R^3} \det(\phi_n, \nabla\phi_n)\,dx\,.$$

If the third alternative (compactness) of the concentration-compactness happened, then up to subsequences, there exists $\,\phi\in X$ such that
\be\label{convg}(\nabla\phi_n, A(\phi_n), \partial_1\phi_n\wedge\partial_2\phi_n\wedge\partial_3\phi_n)\rightharpoonup_n (\nabla\phi, A(\phi), \partial_1\phi\wedge\partial_2\phi\wedge\partial_3\phi)\,,\eq
weakly in $\,L^2(\R^3, S^3)\times L^2(\R^3, S^3)\times L^1(\R^3, S^3)\,$ as $n$ tends to $\,+\infty$.
 Indeed, the convergence of the two first sequences is trivial. As for the last one, note that Proposition \ref{3/4}  makes that sequence locally equi-integrable. On the other hand, the compactness assumption ensures that this sequence is uniformly equi-integrable at infinity. So, up to subsequences, the sequence $\,\{\partial_1\phi_n\wedge\partial_2\phi_n\wedge\partial_3\phi_n\}_n  \,$ is weakly compact in $\,L^1(\R^3, S^3)$. Finally, the particular form of the limits comes from the fact that $\,A(\phi)\,$  and   $\,\partial_1\phi\wedge\partial_2\phi\wedge\partial_3\phi  \,$  are null lagrangians, and hence, they pass to the limit in the weak sense (see the detailed proof in \cite{E2}).
 
 Now, the same kind of arguments show that up to subsequences, 
 $\,\{\det(\phi_n, \nabla\phi_n)\}_n$ is relatively compact in $\,L^1(\R^3, S^3)$-weak. Moreover, from (\ref{convg}) it follows immediately that 
 $$  \det(\phi_n, \nabla\phi_n) \longrightarrow_n \det(\phi, \nabla\phi) \quad\mbox{in}\quad {\mathcal D}'(\R^3)\,. $$
 Hence, up to subsequences, 
  $$  \det(\phi_n, \nabla\phi_n) \longrightarrow_n \det(\phi, \nabla\phi) \quad\mbox{in}\quad L^1(\R^3, S^3)-\mbox{weak} $$
 and so $\,d(\phi)=k\,$ and by lower semicontinuity of the functional $\,{\mathcal E}\,$, $\phi\,$ is a minimizer for $\,I_k$.

Suppose now that dichotomy holds and fix $\epsilon$ small. Let us denote $\,y_n\,$ by $\,y_n^1$, $\,R\,$ by $\,R^1\,$ and $\,R_n\,$ by $R_n^1$.
Let $\bar\phi_1$ be the  weak  limit of $\left\{{\phi_n(\cdot-y_n^1)}\right\}$ in ${\mathcal D}^{1,2}(\R^3)$ (up to extraction of a subsequence). $\,\bar\phi_1\in X$ and
we denote its degree by  $d_1$.  By using the same arguments as those used in the case of compactness, we easily see that up to subsequences,
 $$\frac{1}{2\pi^2}\,\lim_{n\to +\infty}\int_{B(y_n^1, R_n^1)} \det(\phi_n,\nabla\phi_n)\, dx= d_1\,.$$

Now, either $\,d_1=k$ or $\,d_1\neq k$.  If the former happens, then $\,\bar\phi_1\,$ would be a minimizer for $\,I_k$.  Indeed, $\,{\mathcal E}(\bar\phi_1)\leq \liminf_n {\mathcal E}(\phi_n)$. 

If on the contrary $\,d_1\neq k$, let us apply again the concentration-compactness procedure to the sequence $\left\{f_n^2\right\}$, defined as  $\,f_n^2:={f_n}{\large |}_{_{\R^3\setminus B(y^1_n, R_n^1) }}$. We can do it, even if the sequence is not defined in the whole space. 

To start with, there cannot be  vanishing for the sequence $\left\{f_n^2\right\}$.
 Indeed, if this were the case,  $\,d(\phi_n)\,$ would be equal to $\,d_1\,$ for $n$ large enough : if we define  $\,D_n= \R^3\setminus B(y_n^1, R_n^1)\,$,
  \be  \int_{\R^3} \det(\phi_n-P, \nabla \phi_n)\,dx= \int_{D_n}\!\! \det(\phi_n-P, \nabla \phi_n)\,dx +\int_{B(y_n^1, R_n^1)}\!\!\!\!\!\!\!\! \det(\phi_n-P, \nabla \phi_n)\,dx\,,
 \eq
and by Lemma \ref{vanish}, vanishing for the sequence $\,\{f_n^2\}\,$  would imply that \break
$\, \int_{D_n} \det(\phi_n-P, \nabla \phi_n)\,dx\,$ is as small as desired for $n$ large. On the other hand, for $\epsilon>0$ given, let us choose $K_\epsilon>R$ such that
\be\label{degfi}\left|d_1-\int_{B(0, K_\epsilon)} \det(\bar\phi_1-P, \nabla \bar\phi_1)\,dx\right|<\epsilon\,.\eq
Here $R$ stands for the $R$ appearing in the definition of dichotomy. Note that we have used Proposition \ref{degre0} to write (\ref{degfi}).

By choosing $n$ large enough such that $\,K_\epsilon<R_n^1$,
 the dichotomy hypothesis on the sequence $\,\{f_n^1\}\,$ and  $\,\bar\phi_1$'s definition imply that
$$ \lim_{n\to +\infty} \int_{B(y_n^1, R_n^1)} \det(\phi_n-P, \nabla \phi_n)\,dx 
 = O(\epsilon)+\lim_{n\to +\infty} \int_{B(y_n^1, K_\epsilon)} \det(\phi_n-P, \nabla \phi_n)\,dx$$ 
 $$= O(\epsilon)+\int_{B(0, K_\epsilon)} \det(\bar\phi_1-P, \nabla \bar\phi_1)\,dx= O(\epsilon)+d_1\,.$$
Then, taking $\epsilon$ small and $n$ large,  we prove  $\,d(\phi_n)=d_1$, a contradiction. 
So, if $d_1\neq k$, we must again have  either compactness or dichotomy for the restricted sequence $\left\{f_n^2\right\}$ .

Now, as above we find $y^2_n\in \R^3\setminus B(y^1_n, R_n^1)$, $R^2$, $a_2>0$ , $ b_2\geq 0$ and $R^2_n$ going to infinity, such that 
for $n$ large,
$$\left|a_2-\int_{B(y^2_n, R^2)}f^2_n\, dx \right|\leq \epsilon/2, 
\quad \left|\int_{B(y^2_n, R_n^2)\setminus B(y^2_n, R^2)} f^2_n\, dx \right|\leq \epsilon/2\,$$
and
$$\left|b_2-\int_{\R^3\setminus B(y^2_n, R_n^2)}\, f^2_n\, dx  \right|\leq \epsilon/2\,,$$
where $a_2+b_2 = b_1$.

Define $\bar\phi_2$ as the  weak limit of $\left\{{\phi_n(\cdot-y_n^2)}\right\}$ in 
${\mathcal D}^{1,2}(\R^3)$.  We can prove as before that
 $$\frac{1}{2\pi^2}\,\lim_{n\to +\infty}\int_{B(y_n^2, R_n^2)} \det(\phi_n,\nabla\phi_n)\, dx= d(\bar\phi_2)\,.$$

We can iterate this process for every $ j\geq 2$ and find $\,y_n^j\in \R^3\setminus\cup_{i=1}^{j-1}B(y_n^i, R_n^i)\,$, $R^j$, $a_j, \,b_j\geq 0$ and $R_n^j$ going to $\,+\infty$ such that for $n$ large,

$$\left|a_j-\int_{B(y^j_n, R^j)}f^j_n\, dx \right|\leq \epsilon/{2^{j-1}}, 
\quad \left|\int_{B(y^j_n, R_n^j)\setminus B(y^j_n, R^j)} f^j_n\, dx \right|\leq \epsilon/{2^{j-1}}\,$$
and
$$\left|b_j -\int_{\R^3\setminus B(y_n^j, R_n^j)}\, f^j_n\, dx \right|\leq \epsilon/{2^{j-1}}\,,$$
where $a_j+ b_j = b_{j-1}$.

Note that by the definition of dichotomy, the sequence $\,\{a_j\}_j\,$ is nonincreasing. Moreover, 
 for all $\,\ell\geq 1$, $\,\sum_{i=1}^\ell a_j \leq A + 2\epsilon$. Therefore, $\,a_j\,$ tends to $0$ as $j$ tends to $\,+\infty$. Note also that for all $\,j\geq 2$, 
$$ c_n^j:=\sup_{y\in \R^3\setminus\cup_{i=1}^{j-1}B(y_n^i, R_n^i)}\;\int_{B(y,1)  }f_n\,dx\,\leq a_j\longrightarrow_j 0\,.$$

Applying Lemma \ref{vanish} with the choice $\,D_n^j=\R^3\setminus\cup_{i=1}^{j}B(y_n^i, R_n^i)\,$, we infer that for   $\,j\,$ and $\,n\,$  large enough, 
\be\label{JJ}\left|\int_{D_n^j}\det(\phi_n, \nabla\phi_n)\,dx\right|<\pi^2\,.\eq
Let us denote by $J$ the smallest $j$ such that (\ref{JJ}) holds true.

Putting all together we find that
$$k=\sum_{i=1}^J d_j +c_\epsilon+d_n\,,$$
with $\,\lim_{\epsilon\to 0} c_\epsilon=0 $, $ d_n<1/2$ for $n$ large. Since all the $d_j$'s are integers, this means that
$$k=\sum_{i=1}^J d_j\,.$$ 

On the other hand, 
 up to subsequences,
 $$\liminf_{n\to +\infty}\int_{\R^3} \left( |\nabla\phi_n|^2+|A(\phi_n)|^2\right)\,dx\geq \liminf_{n\to +\infty}\Sum_{i= 1} ^J \int_{B(y_n^i, R_n^i)} \left( |\nabla\phi_n|^2+|A(\phi_n)|^2\right)\,dx$$
 $$\qquad\qquad \qquad\qquad\qquad\qquad\geq \Sum_{i= 1}^J\, {\mathcal E}(\bar\phi_i)\geq \Sum_{i= 1}^J\, I_{d_i}\,.$$
 
Hence, since $\,\{\phi_n\}\,$ is a minimizing sequence for $\,I_k$,   we obtain
$$I_k\geq I_{d_1}+\cdots+I_{d_J}\,,$$
with
 $$k=\Sum_{j=1}^{J} d_j\,.$$
 
Therefore, if (\ref{ineq}) holds, dichotomy and vanishing cannot arise and $\,I_k\,$ is attained.\finprf


\begin{thebibliography}{}
%
%

\bibitem{E0} M.J. Esteban. An isoperimetric inequality in $\,\R^3$.  Anal. non lin. Ann.
Inst. H. Poincar\'e 4(4), (1987),  p. 297-305.
\bibitem{E1} M.J. Esteban. A direct variational approach to Skyrme's model for meson fields.
Comm. Math. Phys. 105, (1986), p. 571-591.
\bibitem{E2} M.J. Esteban. A new setting for Skyrme's Problem. Dans "Progress in Nonlinear
Fifferential Equations and Their Applications", vol. 4. Ed. Berestycki et al.,
Birkh\"auser 1990.
\bibitem{E3} M.J. Esteban, S. M\"uller. Sobolev maps with integer degree and applications to Skyrme's problem. Proc. Roy. Soc. London A436 (1992), p. 197-201.
\bibitem{LY1} F. Lin, Y. Yang. Existence of Energy Minimizers as Stable Knotted Solitons in the Faddeev Model. Preprint.
\bibitem{LY2} F. Lin, Y. Yang. Existence of $\,2D\,$ Skyrmions via Concentration-Compactness Method. Preprint.
\bibitem{PLL1} P.-L. Lions. The concentration-compactness
method in the Calculus of Variations. The locally compact
case. Part. I: Anal. non-lin\'eaire, Ann. IHP {\bf 1}
(1984), p. 109-145. Part. II: Anal. non-lin\'eaire, Ann.
IHP {\bf 1} (1984), p. 223-282.


\end{thebibliography}
\end{document}